\documentclass[twocolumn,superscriptaddress,prb]{revtex4-1}

\usepackage{amsmath}
\usepackage{amsfonts}
\usepackage{float}
\usepackage{color}
\usepackage[colorlinks,bookmarks=false,citecolor=blue,linkcolor=red,urlcolor=blue]{hyperref}
\usepackage{ulem}
\setcitestyle{numbers,square}
\usepackage{graphicx,subfigure}

\def \be {\begin{equation}}
	\def \ee {\end{equation}}
\def \ba {\begin{array}}
	\def \ea {\end{array}}
\def \bea {\begin{eqnarray}}
	\def \eea {\end{eqnarray}}

\def \ble {\begin{widetext}\begin{equation}}
		\def \ele {\end{equation}\end{widetext}}
\def \blea {\begin{widetext}\begin{eqnarray}}
		\def \elea {\end{eqnarray}\end{widetext}}

\def \and {{\mathrm{and}}}

\def \T {{\mathrm{T}}}

\begin{document}
	
	
	\title{Quantum Phase Transitions in the Spin-1 Bilinear-Biquadratic Heisenberg Model Based on Classical and Quantum Correlations }

	\author{Ghader Najarbashi}
	
	\affiliation{Department of Physics, University of Mohaghegh Ardabili, P.O. Box 179, Ardabil, Iran}
	
	\author{Hasan Bahmani}
	
	\affiliation{Department of Physics, University of Mohaghegh Ardabili, P.O. Box 179, Ardabil, Iran}
	\author{Babak Tarighi}
	\affiliation{Universidade Federal Fluminense, Niterói, Brazil}

	\begin{abstract}
We investigate thermal and non-thermal quantum correlations in the one-dimensional spin-1 bilinear-biquadratic Heisenberg model. Using tools from quantum information theory—such as generalized concurrence, negativity, and various measures of quantum, classical, and total correlations in bipartite states—we demonstrate that these measures effectively identify quantum phase transitions (QPTs) at critical points. Our negativity analysis reveals nearly identical results at zero or very low temperatures. Importantly, we find that partial concurrence, defined with the reduced density matrix $\rho_{1,2}$, detects more quantum critical points than total concurrence. Additionally, we argue that spin chains with an odd number of spins are more effective than those with an even number in identifying QPTs.	\end{abstract}
	
	\maketitle
	
	
	
	
	
	\section{Introduction}
	Quantum many-body correlated systems have been extensively studied in various physical phenomena, including quantum phase transitions. As a type of quantum critical phenomenon, QPTs occur at absolute zero temperature ($T=0$) and involve essential changes in the ground state of many-body systems when one or more parameters of the Hamiltonian are continuously varied. These qualitative changes are signaled by level crossings in the ground state of quantum many-body systems at a critical value, denoted as $D_c$, of the tuning parameter $D$. At this critical point, the ground state energy becomes non-analytic.
	
    Unlike classical phase transitions, which are driven by thermal fluctuations, QPTs are driven by quantum fluctuations. At or near absolute zero temperature, thermal fluctuations are negligible, and the de Broglie wavelength exceeds the correlation length associated with thermal fluctuations. Examples of QPTs include the paramagnetic-ferromagnetic transition in certain metals, the superconductor-insulator transition, and the superfluid-mott insulator transition \cite {S. Sachdev, M. A. Continentino}.
	\par
	Quantum correlations play a crucial role in quantum information and communication theory \cite{su, se,re}. Typically, quantum correlations can be inferred from entanglement between different system parts \cite{direct, bar}. In recent years, it has been established that there is a general connection between non-analyticities in bipartite entanglement measures and quantum phase transitions \cite{L. Amico, to}. This connection implies that entanglement can signal quantum critical points through the non-analytic behavior inherited from the ground state energy \cite{A. Osterloh, L.-A. Wu, Jaeyoon Cho, ku, tu,qu}.
Various entanglement measures have been identified that effectively describe the special properties of a system near quantum critical points \cite{A. Osterloh, T. J. Osborne, L.-A. Wu, Vidal, Amit Tribedi}. For example, a discontinuity in the first derivative of the ground state energy indicates a first-order phase transition, which corresponds to a discontinuity in a bipartite entanglement measure. Similarly, a discontinuity or divergence in the first derivative of an entanglement measure suggests a second-order phase transition, characterized by a discontinuity or divergence in the second derivative of the ground state energy  \cite{L.-A. Wu}.
	While many studies have focused on the fundamental features of entanglement in spin-$\frac{1}{2}$ models \cite{A. Osterloh, T. J. Osborne, W. K. Wootters, sahar} often have analytical solutions, spin-1 Heisenberg models present a more complex scenario \cite{A. Zheludev, W. Chen, Zheludev, Lou J Z, S.K. Yip,  Fan H, Y. C. Tzeng, Kazuo Hida}. These models exhibit the Haldane gap \cite{Haldane} and a rich phase diagram, highlighting significant differences from their spin-$\frac{1}{2}$ counterparts.
    \par
    To mathematically describe the system at absolute zero temperature, we employ the density matrix formalism. The density matrix $\rho$ for the system at zero temperature is given by:
	
	\begin{equation}
	\rho = |\psi_0 \rangle \langle \psi_0 | 
	\end{equation}
	
	where $|\psi_0 \rangle$ denotes the ground state wavefunction. This density matrix encapsulates all the quantum information of the ground state, enabling the calculation of various physical quantities, such as spin correlations and entanglement entropy.

	At finite temperature $T$, the system is no longer in a pure quantum state but is instead described by a mixed state characterized by a thermal density matrix:
	
	\begin{equation}
	\rho(T) = \frac{e^{-\beta H}}{Z},
	\end{equation}
	
	where $H$ is the Hamiltonian of the system, $Z = \text{Tr}(e^{-\beta H})$ is the partition function, and $\beta = \frac{1}{k_B T}$ represents the inverse temperature. Here, $k_B$ is the Boltzmann constant, which, for simplicity, is often set to unity.
	
	The thermal density matrix $\rho(T)$ captures the effects of thermal fluctuations on the system. The system explores a range of energy states at finite temperatures rather than remaining in a single quantum ground state. The resulting thermal entanglement \cite{Arnesen, Wang, Schliemann}, derived from $\rho(T)$, reflects the interplay between quantum and thermal fluctuations. 
	\par
	In this paper, we employ bipartite entanglement measures—specifically, concurrence\cite{S. Hill, W. K. Wootters, X.-H. Gao} and negativity \cite{G. Vidal} along with measures of quantum, classical, and the total amount of correlations \cite{Tao} in bipartite states, to identify the critical points \cite{H, H1} of quantum phase transitions (QPTs) in the spin-1 bilinear biquadratic chain at both zero and finite temperatures. The paper has three main objectives. First, we aim to investigate the effect of temperature on detecting phase transition points using the negativity measure. To achieve this, we calculate the negativity at zero temperature using the ground state and then evaluate it at finite temperatures using the thermal density matrix. The second objective is to demonstrate that the critical points can be identified by dividing the system into bipartite subsystems. This involves using the reduced density matrix to isolate two particles, which are then separated into bipartite subsystems. For this purpose, we measure entanglement using concurrence as well as Von Neumann entropy to evaluate how effectively this method can detect critical points compared to analyzing the entire bipartite system.

 Additionally, we aim to demonstrate how having an even or odd number of particles in a finite system can influence the results. We use quantum, classical, and total correlation measures to achieve this. Results for open boundary conditions presented in sect.\ref{6} are closely identical to periodic ones.
	
	\section{Hamiltonian}
	
	The Hamiltonian of a one-dimensional spin-1 bilinear biquadratic chain can be expressed as
	
	\begin{equation}
		H = \sum\limits_{i = 1}^{N^{\prime}} \left[\cos\theta \, ({\bf{S}}_i \cdot {\bf{S}}_{i + 1}) + \sin \theta \, ({\bf{S}}_i \cdot {\bf{S}}_{i + 1})^2 \right],
	\end{equation}
	
	where ${{\bf{S}}_i}$ represents the spin-1 operator vector at the $i$-th site of the chain, and $N'$ is the total number of spins in the chain. The parameter $\theta \in [-\pi, \pi)$ determines the relative strength and nature of the bilinear (first term) and biquadratic (second term) interactions between nearest-neighbor spins. Regarding the boundary conditions, when \( N' = N \), the system has periodic boundary conditions, meaning the chain forms a closed loop, so that \( {\bf{S}}_{N + 1} = {\bf{S}}_1 \). In contrast, when \( N' = N - 1 \), the system has open boundary conditions, indicating no correlation between the first and last spins.
	
	In this study, we analyze the behavior of a one-dimensional quantum spin chain, focusing on both its ground state properties at zero temperature ($T = 0$) and its thermal properties at finite temperatures ($T > 0$). At zero temperature, the system is in its ground state, the lowest energy state, where quantum fluctuations are the primary contributors to its behavior. These quantum fluctuations lead to intricate patterns of spin correlations and quantum entanglement, key features of the ground state.
	
    As the temperature rises above zero, thermal fluctuations play a significant role. These fluctuations disrupt the delicate quantum correlations present in the ground state, leading to changes in the spin correlations and the overall entanglement structure of the system. Understanding these changes is crucial for characterizing the thermal properties of the spin chain.

\begin{figure}[H]
        \centering
        \includegraphics[width=0.9\linewidth]{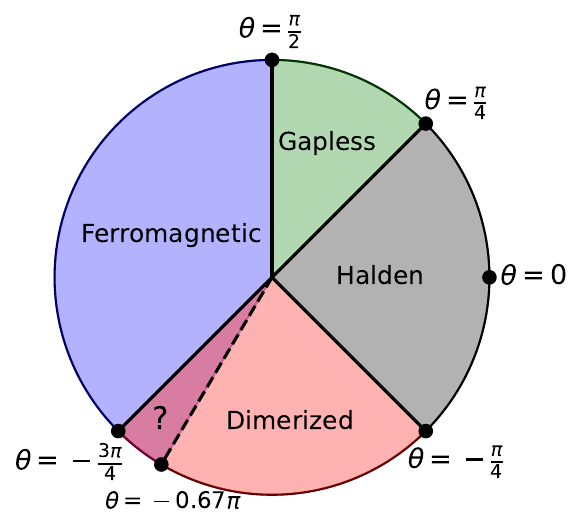}
        \hspace*{0.2cm}
        \caption{\label{Diagram} The phase diagram of the bilinear-biquadratic spin-one chain is shown. The well-established phases include the Haldane phase, the ferromagnetic phase, and the dimerized phase. The extended gapless phase in the range $\frac{\pi}{4} \leq \theta < \frac{\pi}{2}$ is characterized by dominant spin quadrupolar correlations with $k = \pm \frac{2 \pi}{3}$. The potential existence of a spin nematic-like phase near $-\frac{3\pi}{4}$ is explored and critically examined.}
    \end{figure}

    The bilinear-biquadratic spin-one model in one dimension exhibits a rich phase diagram with several distinct regions and transitions, see Figure
    .\ref{Diagram}. For $-\frac{\pi}{4} < \theta < \frac{\pi}{4}$, the system is in the Haldane gap phase, separating a Kosterlitz-Thouless (KT) phase transition from a gapless trimerized phase at $\theta = \frac{\pi}{4}$, where the system also displays $\mathrm{SU}(3)$ symmetry. In the range $\frac{\pi}{4} < \theta < \frac{\pi}{2}$, the system remains in a gapless phase while there is a first-order phase transition from a trimerized phase to the ferromagnetic phase at $\theta = \frac{\pi}{2}$. Another first-order transition occurs at $\theta = -\frac{3\pi}{4}$, moving from the ferromagnetic phase to a gapped dimerized phase. At $\theta = -\frac{\pi}{4}$, the dimerized phase transition changes to the Haldane phase through a second-order transition. Additionally, there is evidence suggesting a non-dimerized nematic phase for $-\frac{3\pi}{4} < \theta < -0.67\pi$ with a KT-type transition at $\theta = -0.67\pi$. Notably, at $\theta = 0.1024\pi$, the system aligns with the Affleck-Kennedy-Lieb-Tasaki (AKLT) model, characterized by an exact valence bond ground state, and at $\theta = -\frac{\pi}{2}$, the model is exactly solvable using the Bethe ansatz method \cite{I. Affleck, M. N. Barber}.

    This study aims to identify the quantum phase transition points described earlier by utilizing various entanglement measures, including negativity, concurrence, and correlations. We investigate how these measures can effectively detect transitions by examining their behavior across different parameters, such as temperature, the size of the system (both the entire system and its reduced subsystems), and the number of particles involved. By analyzing these factors, we seek to understand how they influence the ability of entanglement measures to reveal critical points and provide insights into the underlying quantum phases.
	
	\section{Effect of Temperature on Transition Detection}
	Negativity stems from the Peres-Horodecki separability criterion, which provides a method for determining the entanglement of a bipartite quantum state \cite{A. Peres, M}. According to this criterion, a bipartite state $\rho_{AB}$ in the Hilbert space $H_A \otimes H_B$ is identified as entangled if the partial transpose of the state concerning one of its subsystems has any negative eigenvalues. Specifically, the partial transpose of $\rho_{AB}$ with respect to subsystem $B$ can reveal the presence of entanglement through its eigenvalue spectrum. The state is confirmed to be entangled if any of these eigenvalues are negative.

    Negativity, $\mathcal{N}$, quantifies the extent of entanglement and is defined as the sum of the absolute values of the negative eigenvalues of $\rho_{AB}^{T_B}$
    
    \begin{equation}
    \mathcal{N}(\rho_{AB}) = \sum\limits_i \left| \mu_i \right|,
    \end{equation}
    
    where $\mu_i$ are the negative eigenvalues of the partially transposed density matrix $\rho_{AB}^{T_B}$. The partial transpose operation $T_B$ is applied to subsystem $B$ of the bipartite system. The negativity thus serves as a quantitative measure derived from the Peres-Horodecki criterion, allowing for assessing the degree of entanglement in the system.
    
    If $\mathcal{N} > 0$, it indicates that the state is entangled and exhibits quantum correlations that cannot be explained by classical separability alone. Therefore, negativity is a practical and widely used tool for quantifying entanglement in bipartite quantum systems, providing critical insights into the nature of quantum correlations.

	\par
\begin{figure}[H]
        \centering
        \includegraphics[width=1.1\linewidth]{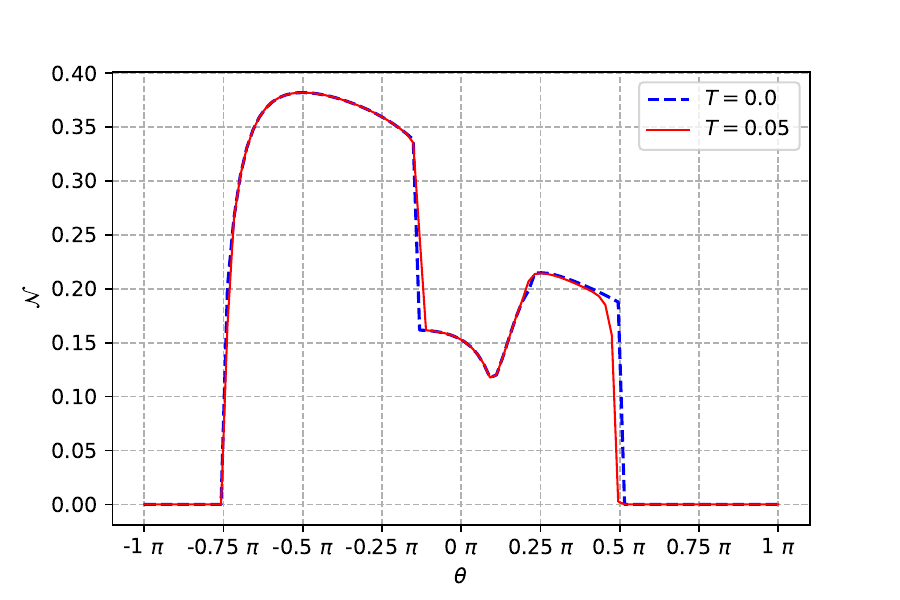}
        \hspace*{0.2cm}
        \caption{\label{JPT1} Negativity as a function of angle \(\theta\) for a quantum spin chain of size \(L=7\) and with periodic boundary condition, at ground state \(T=0\) (blue dash line) and thermal state \(T=0.05\) (red solid line). The sharp transitions at specific \(\theta\) values signal quantum phase transitions, with first-order transitions occurring at \(\theta = -0.75\pi\), \(\theta = 0.1\pi\), and \(\theta = 0.5\pi\), and second-order transitions at \(\theta = -0.1\pi\) and \(\theta = 0.25\pi\). The close similarity in negativity between \(T=0\) and \(T=0.05\) indicates the robustness of entanglement against low thermal fluctuations. The system exhibits distinct entanglement structures in different quantum phases, including antiferromagnetic, ferromagnetic, and trimerized phases.}
    \end{figure}

	Figure \ref{JPT1} illustrates the behavior of negativity as a function of the angle $\theta$ for a system size of $L=7$, comparing the ground state at $T=0$ and the thermal state at $T=0.05$. The figure reveals that sharp transitions in negativity occur at specific values of $\theta$: namely, $\theta = -0.75\pi$, $\theta = -0.1\pi$, $\theta = 0.1\pi$, $\theta = 0.25\pi$, and $\theta = 0.5\pi$. These transitions indicate significant changes in the entanglement structure of the system, highlighting the sensitivity of negativity to variations in $\theta$ at both zero and finite temperatures.

	\par
	Within the interval $-0.25\pi < \theta < 0.25\pi$, a Haldane gap exists between a spin-singlet ground state and a spin-triplet excited state, a characteristic feature of spin-$1$ chains. The figure demonstrates that the system undergoes first-order quantum phase transitions (QPTs) at $\theta = -0.75\pi$, $\theta = 0.1\pi$, and $\theta = 0.5\pi$. In contrast, at $\theta = -0.1\pi$ and $\theta = 0.25\pi$, the system exhibits second-order QPTs.
	
	The results for the negativity measure are nearly identical at $T=0$ and $T=0.05$, indicating that the temperature has little impact when it is either zero or very low. This suggests that the system's entanglement properties, as captured by negativity, are robust against small thermal fluctuations in this temperature range.
    \par
    The system resides in the antiferromagnetic phase for $- \frac{3}{4}\pi < \theta < \frac{\pi}{2}$, while it is in the ferromagnetic phase for the remaining range of $\theta$. Within the interval $ \frac{\pi}{4} < \theta < \frac{\pi }{2}$, the system enters a trimerized phase, characterized by a gapless spectrum between the spin-triplet ground state and a fivefold degenerate excited state. A phase transition occurs at the critical point $\theta = 0.25\pi$, marking the boundary between the Haldane and trimerized phases. This transition signifies a change in the system's ground state structure and its associated entanglement properties \cite{Y Xian}.

	\section{Partial and Total States in Transition Detection}\label{sec:Concurrence}
	Concurrence, introduced by Hill and Wootters, is an entanglement measure derived from the Entanglement of Formation (EoF) to quantify the entanglement in pure two-qubit states. Because the EoF is a monotonically increasing function of concurrence, concurrence serves effectively as a measure of entanglement. It is beneficial for assessing entanglement in spin-$\frac{1}{2}$ systems, with values ranging from 0 for separable states to 1 for maximally entangled Bell states. Wootters showed that the Entanglement of Formation for a two-qubit mixed state $\rho$ is related to the concurrence $C(\rho)$ by:

    \begin{equation}
        E_F(\rho) = H\left( \frac{1}{2} + \frac{1}{2} \sqrt{1 - C^2} \right),
    \end{equation}
    
    where $H(x)$ is the binary entropy function defined as
    
    \begin{equation}
        H(x) = -x \ln x - (1 - x) \ln (1 - x),
    \end{equation}
    
    and $C(\rho)$ represents the concurrence, which is given by
    
    \begin{equation}
        C(\rho) = \max \{ 0, \lambda_1 - \lambda_2 - \lambda_3 - \lambda_4 \}.
    \end{equation}
    
    Here, $\lambda_1, \lambda_2, \lambda_3, \lambda_4$ are the square roots of the eigenvalues of the matrix $R = \rho (\sigma_y \otimes \sigma_y) \rho^* (\sigma_y \otimes \sigma_y)$, listed in descending order, where $\rho^*$ is the complex conjugate of $\rho$.
 
    \par
    Recently, it has been demonstrated that concurrence is essential for detecting critical points of quantum phase transitions (QPTs) in various interacting quantum many-body systems. While most studies focus on spin-$\frac{1}{2}$ Heisenberg chains, entanglement properties in spin-1 chains are less explored due to the lack of effective operational measures for higher spin systems. Li et al. extended the Hill-Wootters concurrence to qutrits and higher-dimensional systems, as described in Ref. \cite{Y. Q. Li}. We utilize this generalized concurrence vector, where its norm can be used to quantify the entanglement of both pure and mixed states. The generalized concurrence measure introduced in Ref. \cite{Y. Q. Li} is defined as:

    \begin{equation}
        \left| \mathbf{C} \right|^2 = \sum_{\alpha \beta} C_{\alpha \beta}^2,
    \end{equation}
    
    where $C_{\alpha \beta}$ are the components of the concurrence vector $\mathbf{C}$, given by:
    
    \begin{equation}
        C_{\alpha \beta}(\rho) = \max \left\{ 0, 2 \max (\lambda_{i}^{\alpha \beta}) - \sum_{i} \lambda_{i}^{\alpha \beta} \right\},
    \end{equation}
    
    where, the values $\lambda_{i}^{\alpha \beta}$, where $i = 1, \ldots, 4$, denote the square roots of the eigenvalues of the matrix $\rho \left( L_{\alpha} \otimes L_{\beta} \right) \rho^* \left( L_{\alpha} \otimes L_{\beta} \right)$. In this context, $L_{\alpha}$ and $L_{\beta}$ are the generators of the special orthogonal groups $SO(d_1)$ and $SO(d_2)$, respectively. Specifically, $L_{\alpha}$ corresponds to the generators for $\alpha$ ranging from 1 to $\frac{d_1(d_1 - 1)}{2}$, and $L_{\beta}$ corresponds to the generators for $\beta$ ranging from 1 to $\frac{d_2(d_2 - 1)}{2}$. These generators are crucial for constructing the matrix used in the calculation of concurrence, as they help to define the tensor product spaces involved. The matrix $\rho \left( L_{\alpha} \otimes L_{\beta} \right) \rho^* \left( L_{\alpha} \otimes L_{\beta} \right)$ is formed by applying the tensor product of these generators to the density matrix $\rho$ and its complex conjugate $\rho^*$, and the resulting eigenvalues provide the necessary components to compute the concurrence.

	\par
\begin{figure}[H]
        \centering
        \includegraphics[width=1.1\linewidth]{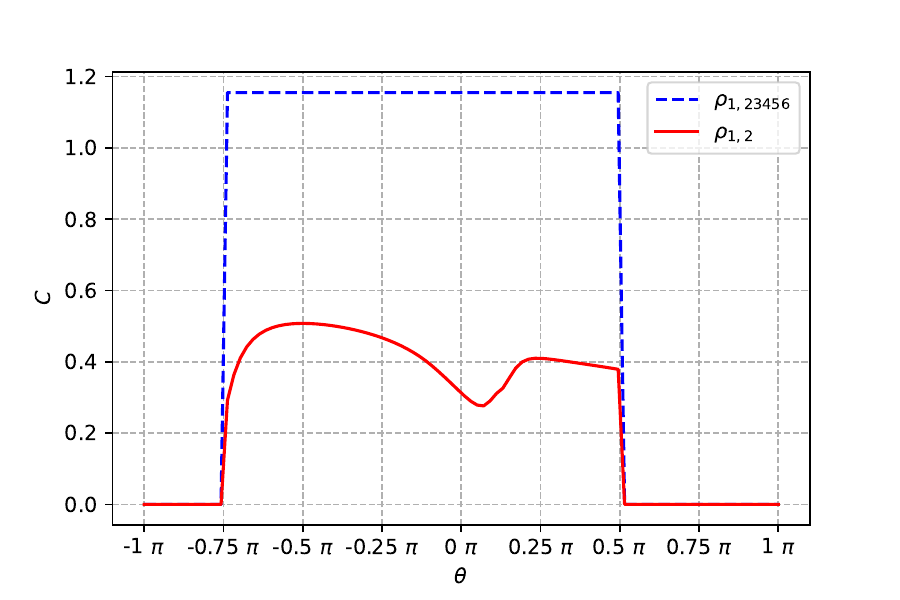}
        \hspace*{0.2cm}
        \caption{\label{JPT2} Concurrence as a function of the angle \(\theta\) is shown for a quantum spin chain of size \(L=6\) and with periodic boundary condition, comparing the entire state \(\rho_{1,23456}\) (blue dashed line) and a partial state \(\rho_{1,2}\) (red solid line). The sharp changes in concurrence at specific values of \(\theta\) indicate quantum phase transitions, with first-order transitions observed at \(\theta = -0.75\pi\), \(\theta \approx 0.1024\pi\), and \(\theta = 0.5\pi\). The graph demonstrates that the partial state \(\rho_{1,2}\) identifies an additional transition point (\(\theta \approx 0.1024\pi\)) compared to the entire system, despite requiring fewer computational resources.}
    \end{figure}

	The variation of total concurrence and partial concurrence as a function of \(\theta\) for a quantum spin chain of length \(L = 6\) at \(T = 0\) is shown in the figure (see Figure \ref{JPT2}). The total concurrence, represented by the dashed line, measures the overall entanglement of the entire system between the first site and the rest of the system $(\rho_{1,23456})$. In contrast, partial concurrence, shown by the solid line, quantifies the entanglement between specific subsets of spins, particularly those located at sites 1 and 2, while tracing the degrees of freedom associated with the other spins. Mathematically, this is represented by the reduced density matrix \(\rho_{1,2}\), which is derived by tracing out all spins except for those at sites 1 and 2.

    Within the range of \(-0.75\pi \leq \theta \leq 0.5\pi\), the figure demonstrates that the total concurrence is consistently greater than the partial concurrence, indicating a more substantial overall entanglement in this interval. Both entanglement measures vanish outside this range, suggesting that the system is in a separable state or a phase where entanglement is absent. The distinction between total and partial concurrence becomes particularly relevant when examining quantum phase transitions. The partial concurrence is often more sensitive to these transitions; for example, around \(\theta \approx 0.1024\pi\), it detects a critical point associated with the Affleck-Kennedy-Lieb-Tasaki (AKLT) model, which signifies a quantum phase transition that is not as clearly identified by the total concurrence. This sensitivity to different phases and transitions highlights the usefulness of partial concurrence in studying spin-1 chains, where localized entanglement properties can provide deeper insights into the underlying quantum phenomena.

    As an additional measure, we can examine the von Neumann entanglement entropy,$S(\rho)$, to identify the phase transition points.  For the pure quantum states and a bipartition, the von Neumann entanglement entropy, also known as the entropy of entanglement or just the entanglement entropy of subsystem $A$ through the Schmidt decomposition is computed as follows 
    \begin{equation}\label{von_Neumann1}
    S(\rho_A) = -\sum_{i=1}^N \lambda_i^2 \ln \lambda_i^2,
    \end{equation} 
    where $\lambda_i$ is Schmidt coefficients.
    Alternatively, employing a density matrix as a more comprehensive means of describing a quantum system allows us to establish von Neumann entropy through a reduced density matrix. For a pure state $\rho_{AB}=|\Psi \rangle \langle \Psi |_{AB}$, it is given by:
    \begin{equation}\label{von_Neumann2}
        S(\rho_A) = -\mathrm{Tr}[\rho_A \ln(\rho_A)],
    \end{equation}
    where $\rho_{A} = \operatorname{Tr}_{B}(\rho_{AB})$ is the reduced density matrix for the subsystem $A$ and $S(\rho_A)$ is the amount of entangelement entropy between subsytems $A$ and $B$. Since both equations \ref{von_Neumann1} and \ref{von_Neumann2} represent the von Neumann entropy, it follows that \(\lambda_i\) are the eigenvalues of \(\rho_A\). 
    Notably, the value of von Neumann entropy remains consistent, regardless of the choice of the subsystem, whether it's A or B. Thus
    \begin{equation}
        \mathcal{S}(\rho_{A}) = -\operatorname{Tr}[\rho_{A}\ln \rho_{A}] = -\operatorname{Tr}[\rho_{B}\ln \rho_{B}] = \mathcal{S}(\rho_{B}),
    \end{equation}
    where $\rho_{B} = \operatorname{Tr}_{A}(\rho_{AB})$ are the reduced density matrices for partition $B$.

    \begin{figure}[H]
        \centering
        \includegraphics[width=1.1\linewidth]{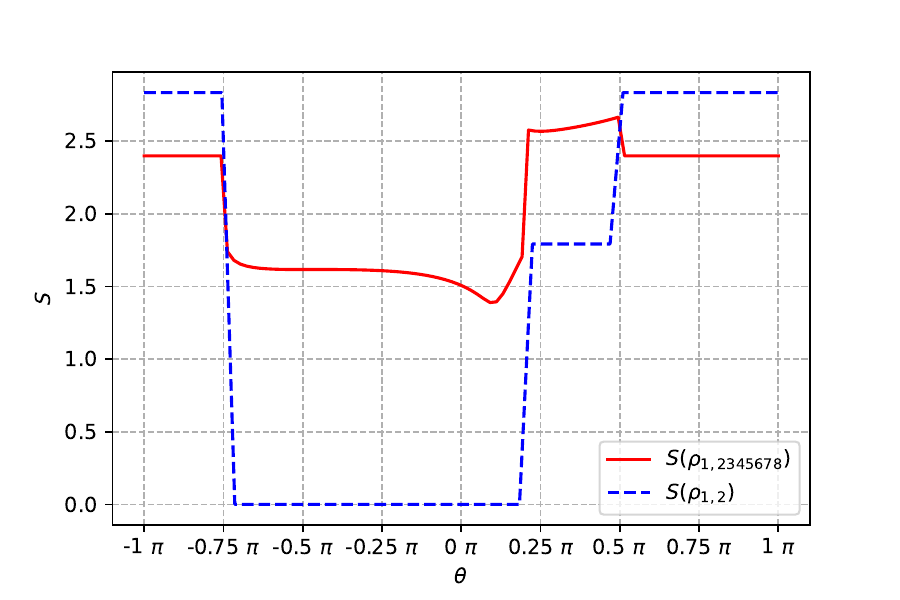}
        \hspace*{0.2cm}
        \caption{\label{JPT5} Von Neumann entropy for the total state (red solid line) and the partial state (blue dashed line) is shown for a system with size \(L = 8\) and periodic boundary condition. The entanglement entropy measure for the total state detects transition points at \(\theta = -0.75\pi\), \(\theta = 0.5\pi\), \(\theta \approx 0.1024\pi\) and \(\theta = 0.25\pi\). The von Neumann entropy for the partial state, despite being a mixed state and not a direct measure of entanglement, also identifies transition points at \(\theta = -0.75\pi\), \(\theta = 0.5\pi\), and \(\theta = 0.25\pi\).}
    \end{figure}

    Figure \ref{JPT5} shows the results for a system size of \(L = 8\). The calculation considers the entanglement between site 1 and the system's remaining part, consisting of 7 spin-1 particles. The red solid line represents this. This measure effectively detects phase transition points at \(\theta = -0.75\pi\) and \(\theta = 0.5\pi\), which aligns with the results obtained using the concurrence for the entire system. However, the von Neumann entanglement entropy reveals two additional transition points at \(\theta \approx 0.1024\pi\) and \(\theta = 0.25\pi\). Moreover, the von Neumann entropy, when applied to the partial state, which itself is a mixed state and thus not a direct measure of entanglement, still identifies transition points at \(\theta = -0.75\pi\), \(\theta = 0.5\pi\), and \(\theta = 0.25\pi\). This suggests that the von Neumann entropy is sensitive to the entanglement properties of the total state and critical behavior in mixed states.

    \section{Even and Odd Particles in Transition Detection}
    This section introduces a comprehensive measure to quantify quantum, classical, and total correlations in bipartite quantum states. This measure was originally developed by Zhou et al. in \cite{T. Zhou}, and it is based on a necessary and sufficient condition for identifying zero-discord states. Discord is a well-known measure in quantum information theory that quantifies the amount of non-classical correlations present in a quantum system, even beyond entanglement. A general bipartite quantum state ${\rho_{AB}}$ defined on the Hilbert space ${H_{AB}}$ can be expressed using the coherence vector representation as follows:
        \begin{equation}
		\begin{array}{l}
			{\rho _{AB}} = \frac{1}{{{d_A}{d_B}}}{I_A} \otimes {I_B} + \frac{1}{{2{d_B}}}\sum\limits_{i = 1}^{d_A^2 - 1} {{\lambda _{Ai}}\left( {{{\hat \lambda }_{Ai}} \otimes {I_B}} \right)} \\
			+ \frac{1}{{2{d_A}}}\sum\limits_{i = 1}^{d_B^2 - 1} {{\lambda _{Bi}}\left( {{I_A} \otimes {{\hat \lambda }_{Bi}}} \right)}  + \frac{1}{4}\sum\limits_{i = 1}^{d_A^2 - 1} {\sum\limits_{j = 1}^{d_B^2 - 1} {{K_{ij}}\left( {{{\hat \lambda }_{Ai}} \otimes {{\hat \lambda }_{Bj}}} \right)} } 
		\end{array}
	\end{equation} 
        \\
        In this expression, the matrices $\{ {\lambda _{{A_i}}}:i = 1,2,...,d_A^2 - 1\}$ and $\{ {\lambda _{{B_i}}}:i = 1,2,...,d_B^2 - 1\}$ represent the coherence vectors associated with the reduced density matrices ${\rho _A}$ and ${\rho _B}$ of subsystems A and B, respectively. The identity matrix $I$ represents the identity operation over the corresponding subsystem, ensuring the proper normalization of the density matrix. The local Bloch vectors ${\lambda _A}$ and ${\lambda _B}$, as well as the second-order correlation tensor ${{\bf{K}}_{ij}}$, are defined to capture the degrees of correlation between different parts of the bipartite state. The specific components of these vectors and tensors are given by:
        
        \begin{equation}
		\begin{array}{l}
			{\lambda _{Ai}} = tr\left( {{\rho _{AB}}{{\hat \lambda }_{Ai}} \otimes {I_B}} \right),\\
			{\lambda _{Bi}} = tr\left( {{\rho _{AB}}{I_A} \otimes {{\hat \lambda }_{Bi}}} \right),\\
			{K_{ij}} = tr\left( {{\rho _{AB}}{{\hat \lambda }_{Ai}} \otimes {{\hat \lambda }_{Bj}}} \right),
		\end{array}
	\end{equation}
        
        Here, the generators ${\hat \lambda _{Ai}}$ and ${\hat \lambda _{Bi}}$ are chosen such that they fulfill specific algebraic properties: 
        
        \begin{equation}
		\hat \lambda _i^\dag  = {\hat \lambda _i},\,\,\,\,\,\,\,\,\,\,\,tr\left( {{{\hat \lambda }_i}} \right) = 0,\,\,\,\,\,\,\,\,\,\,\,tr\left( {{{\hat \lambda }_i}{{\hat \lambda }_j}} \right) = 2{\delta _{ij}}.\,
	\end{equation}
        
        These properties ensure that the generators form an orthonormal basis set under the trace inner product, crucial for defining the state's geometry in the associated Hilbert space. 
        
        Utilizing the structure of bipartite quantum states and the formalism described in theorem 2 of Ref \cite{T. Zhou}, we derive measures for the quantum, classical, and total correlations within the state:
        
        \begin{equation}
		\begin{aligned}
			\tau \left( \rho_{AB} \right) &= \frac{1}{4}\sum_{i=1}^{d_A^2-1} \left| \Lambda_i \right|, \\
			C\left( \rho_{AB} \right) &= \frac{1}{4}\sum_{i=1}^{d_A-1} \left| \Lambda_i \right|, \\
			Q\left( \rho_{AB} \right) &= \frac{1}{4}\sum_{i=d_A}^{d_A^2-1} \left| \Lambda_i \right|.
		\end{aligned}
		\label{eq:equation_label}
	\end{equation}
        
        where ${\Lambda_i}$ are obtained by diagonalizing the criterion matrix ${\bf{\Lambda }} = {\bf{K}}{{\bf{K}}^T} - {\bf{\lambda }}_B^2{{\bf{\lambda }}_A}{\bf{\lambda }}_A^T$. These measures allow us to quantify different types of correlations within a bipartite quantum system, providing deeper insight into the interplay between quantum and classical phenomena at a fundamental level.

	\par
\begin{figure}[H]
        \centering
        \includegraphics[width=1.1\linewidth]{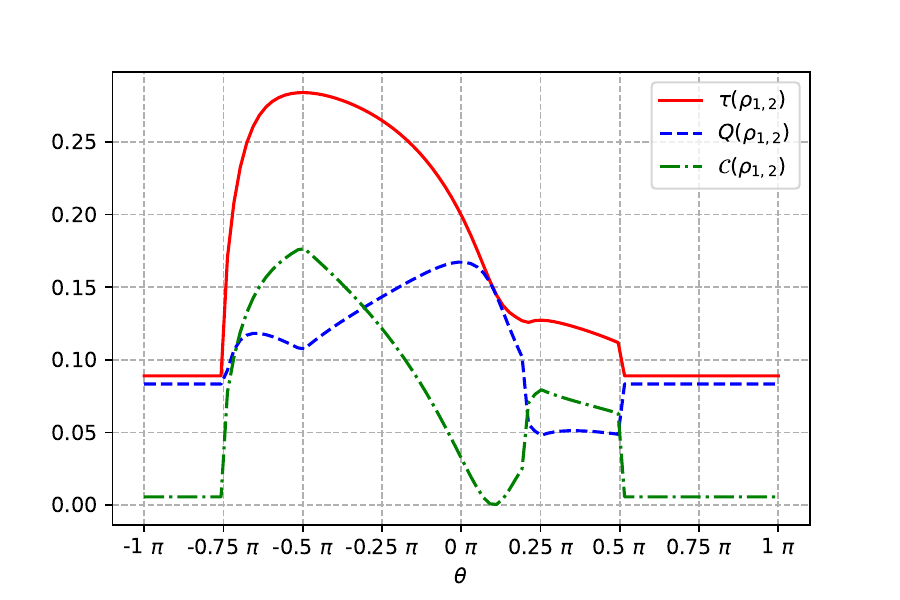}
        \hspace*{0.2cm}
        \caption{\label{JPT3} Variation of total correlations (solid red line), classical correlations (dash-dot green line), and quantum correlations (dash blue line) as a function of $\theta$ for a spin chain with $L=8$ at $T=0$. Classical and quantum correlations converge at the quantum phase transition points: $\theta = -0.75\pi$, $\theta = -0.5\pi$, $\theta \approx 0.1024\pi$, $\theta = 0.25\pi$, and $\theta = 0.5\pi$. However, the total correlations only signal quantum critical points at $\theta = -0.75\pi$ and $\theta = 0.5\pi$.}
    \end{figure}

 
	Figure \ref{JPT3} shows the classical, non-classical, and total correlations for a system with $L=8$ at $\T=0$. In this analysis, the reduced density matrix ${\rho _{1,2}}$ is obtained by taking the partial trace of the system's density operator over all spins except those at sites 1 and 2. The figure shows that classical and non-classical correlations align precisely at the critical points of the quantum phase transition. These critical points occur at $\theta=-0.75\pi$, $\theta=-0.5\pi$, $\theta\approx0.1024\pi$, $\theta=0.25\pi$, and $\theta=0.5\pi$. However, the total correlations only indicate the presence of two quantum critical points at $\theta=-0.75\pi$ and $\theta=0.5\pi$.

\begin{figure}[H]
		\centering
		\includegraphics[width=1.1\linewidth]{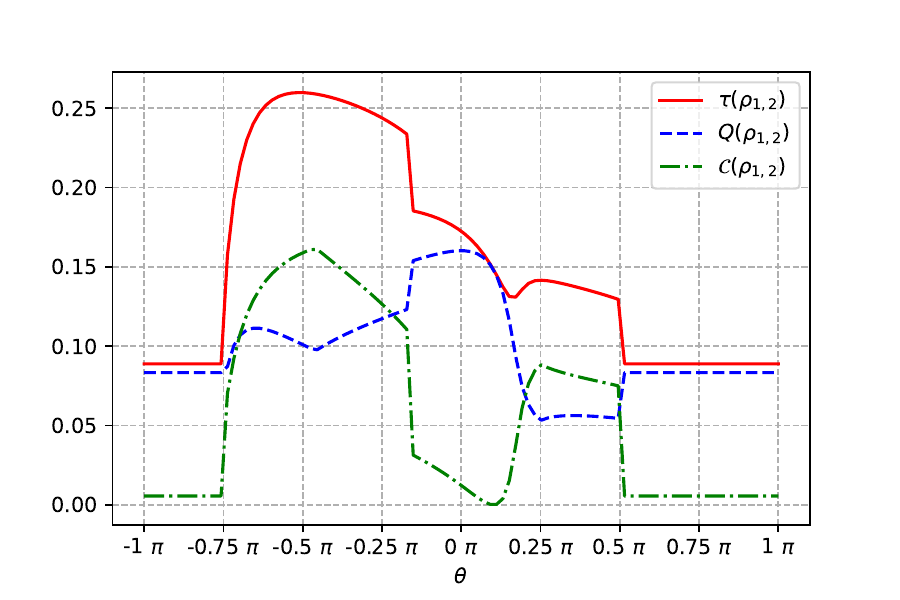}
		\hspace*{0.2cm}
		\caption{\label{JPT4} Variation of total correlations (solid red line), classical correlations (dash-dot green line), and quantum correlations (dash blue line) as a function of $\theta$ for a spin chain with $L=9$ at $T=0$. The total correlations show an additional quantum critical point compared to the $L=8$ case. A sharp transition is observed in both quantum and classical correlations at $\theta = -0.22\pi$, close to the expected transition point at $\theta = -0.25\pi$. This suggests that this transition point may converge to $\theta \approx -0.25\pi$ as the system size increases.}
    \end{figure}

    Figure \ref{JPT4} displays the correlation data for a system with $L=9$. For this case, the total correlations reveal an additional quantum critical point compared to the $L=8$ case. Additionally, there is a distinct sharp transition in both quantum and classical correlations at $\theta=-0.22\pi$, which is near the expected transition point at $\theta=-0.25\pi$. Observing the behavior of smaller systems with an odd number of spins suggests that this transition point will likely converge to $\theta \approx -0.25\pi$ as the system size increases.

    These findings indicate that for finite-sized spin chains, systems with an odd number of spins (such as $L=9$) generally exhibit more quantum critical points than those with an even number of spins (such as $L=8$). However, as the system size increases toward infinity, the specific distinction between even and odd numbers of spins becomes less meaningful. In the thermodynamic limit (infinite system size), the differences observed due to parity (even vs. odd) are expected to diminish, as the system's bulk properties dominate over any finite-size or boundary effects.

    Despite this, studying finite systems provides valuable insights into the nature of quantum phase transitions. The additional quantum critical points observed in finite systems with an odd number of spins suggest that these systems may have more complex correlations and richer phase structures. As the system approaches infinite size, these observations help us predict how the phase transitions might manifest, even though the number of particles' exact parity (even or odd) will no longer be relevant. Instead, what remains important is understanding the general scaling behavior and how quantum correlations evolve as the system grows.

\begin{figure}[H]
        \centering
        \includegraphics[width=1.1\linewidth]{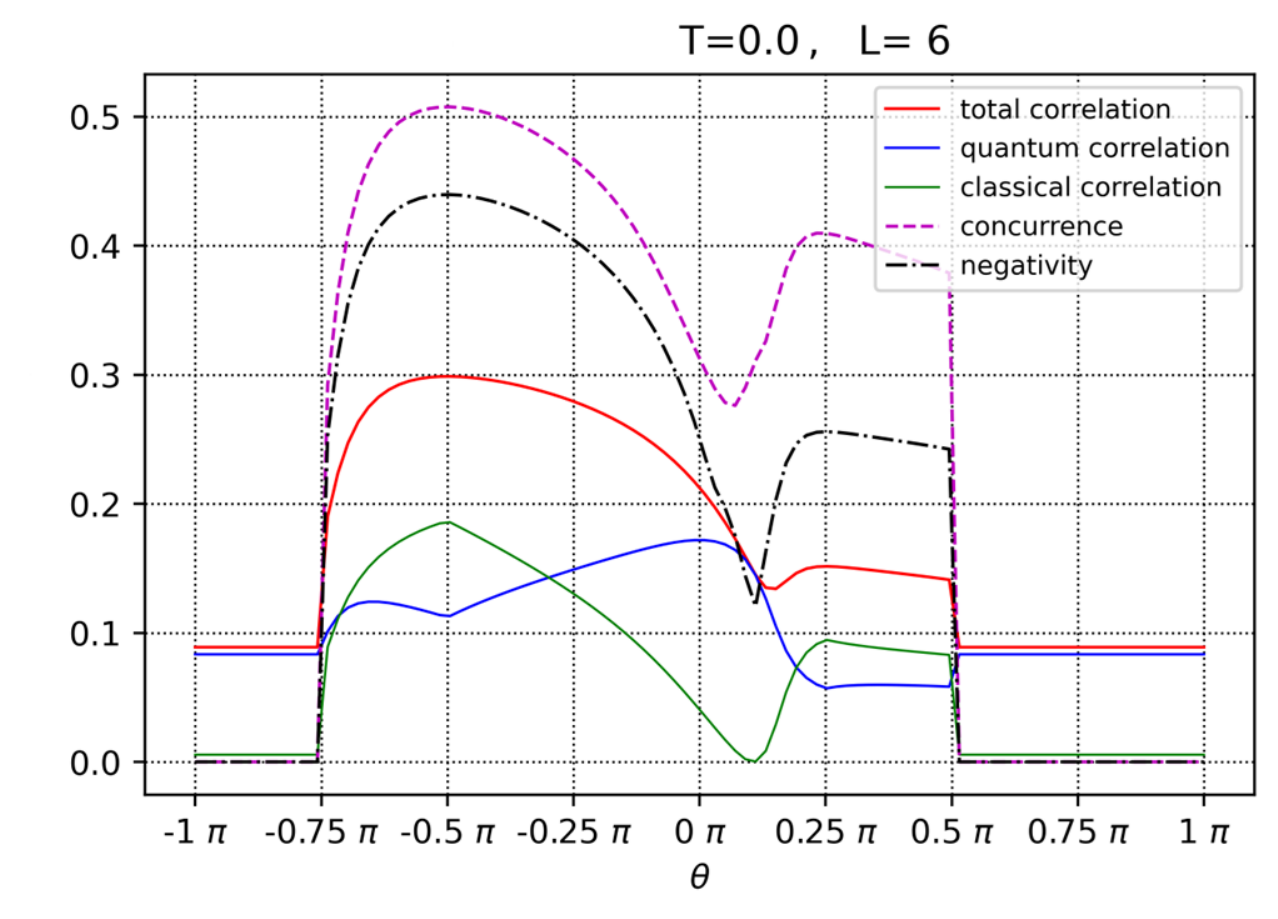}
        \hspace*{0.2cm}
        \caption{\label{all_methods} All methods in one figure for $L=6$ at $T=0$.}
    \end{figure}

    As a summarization of all methods, we can plot all methods in one figure for $L=6$ at $T=0$ in the figure \ref{all_methods}.

    \section{Analysis of Boundary Conditions: Open vs. Periodic Systems}\label{6}
\begin{figure}[H]
		\centering
		\includegraphics[width=1.1\linewidth]{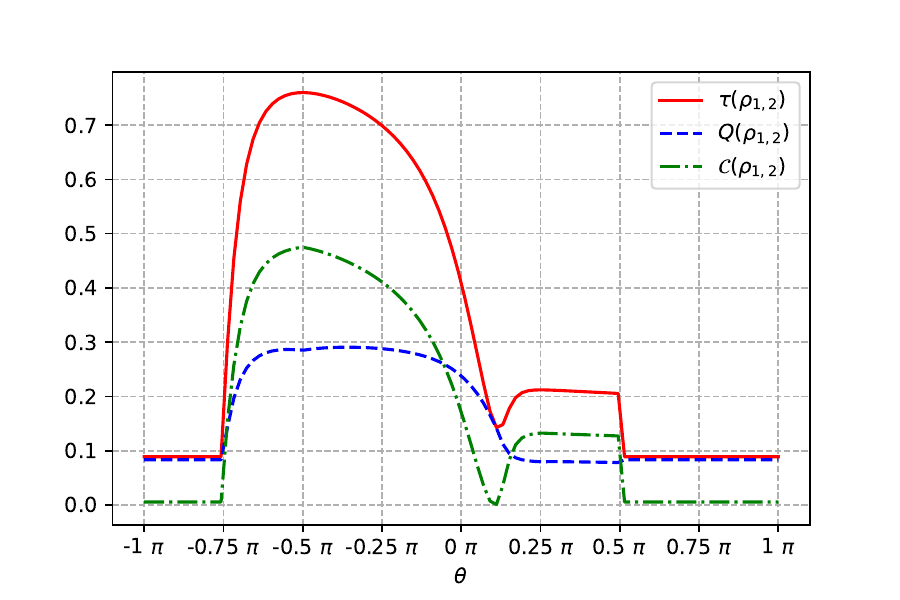}
		\hspace*{0.2cm}
		\caption{\label{JPT6} Total (solid red), classical (dash-dot green), and quantum (dash blue) correlations vary with $\theta$ for an $L=9$ spin chain with open boundaries at $T=0$. Compared to the $L=8$ case, total correlations reveal an additional quantum critical point. A sharp transition in quantum and classical correlations occurs near $\theta = -0.22\pi$, suggesting this point may converge to $\theta \approx -0.25\pi$ as system size increases.}
    \end{figure}

    The distinction between systems with periodic and open boundary conditions lies in the interaction between the first and last particles, present in periodic boundaries but absent in open ones. To identify phase transition points, we analyze the classical, quantum, and total correlations for systems with both even and odd particle numbers under open boundary conditions. These results are compared with Figures \ref{JPT3} and \ref{JPT4}, which represent systems with periodic boundary conditions. Figures \ref{JPT5} and \ref{JPT6} demonstrate that, in general, the transition points identified in open boundary systems closely align with those in periodic systems, except for an additional transition point detected in the latter. However, the transition points are sometimes less distinct under open boundary conditions. This suggests that increasing the particle count could enhance the clarity of these transitions in open boundary systems, making them comparable to those observed with periodic boundaries.

    \begin{figure}[H]
        \centering
        \includegraphics[width=1.1\linewidth]{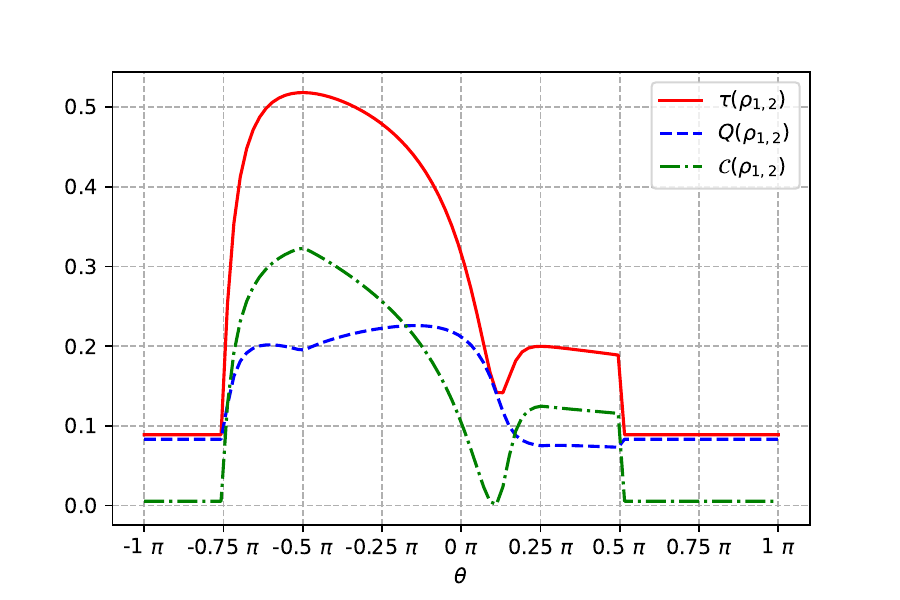}
        \hspace*{0.2cm}
        \caption{\label{JPT5} Total correlations (solid red), classical correlations (dash-dot green), and quantum correlations (dash blue) vary with $\theta$ for an $L=8$ spin chain with open boundaries at $T=0$. Classical and quantum correlations meet at quantum phase transition points: $\theta = -0.75\pi$, $\theta = -0.5\pi$, $\theta \approx 0.1024\pi$, $\theta = 0.25\pi$, and $\theta = 0.5\pi$. Total correlations, however, indicate quantum criticality only at $\theta = -0.75\pi$ and $\theta = 0.5\pi$.}
    \end{figure}

    The additional transition point detected in systems with periodic boundaries arises from the symmetry described by the \( Z_n \) group. The \( Z_n \) group, or cyclic group of order \( n \), consists of \( n \) elements \( \{0, 1, \dots, n-1\} \) with addition modulo \( n \) as its operation. This group is abelian (commutative) and cyclic, meaning all elements can be generated from a single element, such as \( 1 \). Geometrically, \( Z_n \) represents the rotational symmetries of a regular \( n \)-gon. Its symmetry, which originates from the interaction between the first and last particles, is absent in systems with open boundaries.

    Due to the \( Z_n \) group symmetry inherent in periodic boundary systems, by the definition, the measures detect one additional transition point compared to open boundary systems. This highlights the role of boundary conditions in shaping the behavior of phase transitions.

    \section{Conclusion}

    In conclusion, we investigated the quantum phase transitions (QPTs) and their detection through various entanglement measures, focusing on the one-dimensional spin-1 bilinear-biquadratic Heisenberg model. Our findings reveal that negativity, concurrence, and von Neumann entropy provide critical insights into phase transitions at zero and finite temperatures.
    
    Negativity was shown to detect first effectively- and second-order phase transitions, exhibiting robustness even at low temperatures. Concurrence, particularly when applied to partial subsystems, demonstrated the ability to identify additional quantum critical points that are not as evident in the analysis of the entire system. This highlights the advantage of using localized entanglement measures in detecting more subtle transitions, such as those associated with the Affleck-Kennedy-Lieb-Tasaki (AKLT) model.
    
    Moreover, the study of systems with both even and odd numbers of particles underscored a key observation: odd-numbered spin chains generally identified more critical points than even-numbered chains. This suggests that spin parity plays a significant role in phase detection in finite systems, although this effect diminishes as the system approaches the thermodynamic limit.
    
    Overall, our results demonstrate that entanglement measures are indispensable for exploring QPTs. They provide a quantitative means of detecting critical behavior and offer deeper insights into the interplay between quantum and classical correlations, particularly in spin-1 Heisenberg models. As such, these findings can contribute to a broader understanding of quantum many-body systems and their phase structures. Future research could explore how these methods extend to higher-dimensional systems and other spin models.


\begin{thebibliography}{99}
\bibitem{S. Sachdev}
S. Sachdev,
  \textit{{Quantum Phase Transitions}},
  \href{https://doi.org/10.1017/CBO9780511973765}  {Cambridge Univ. Press (2011).}

\bibitem{M. A. Continentino}
M. A. Continentino, 
  \textit{{Quantum Scaling in Many-Body Systems}},
  \href{https://doi.org/10.1017/CBO9781316576854} {Cambridge Univ. Press. (2017).} 
  
  
 \bibitem{su}
 YC. Li, HQ Lin
\textit{{Quantum coherence and quantum phase transition}},
\href{https://doi.org/10.1038/srep26365/Scientific Reports volume 6}{Sci Rep {\bfseries 6}, 26365 (2016)}.
  
\bibitem{se}
Z. Wang , Q. Zheng,  X. Wang et al
\textit{{The energy-level crossing behavior and quantum Fisher information in a quantum well with spin-orbit coupling}},
\href{https://doi.org/10.1038/srep22347/Scientific Reports volume 6}{Sci Rep {\bfseries 6}, 22347 (2016)},   [\href{https:// arXiv:quant-ph/1510.06491 }{{\ttfamily
 			arXiv:quant-ph/1510.06491}}].	


\bibitem{re}
 X. Liu, W. Cheng, JM. Liu
\textit{{Renormalization-group approach to quantum Fisher information in an XY model with staggered Dzyaloshinskii-Moriya interaction}},
\href{https://doi.org/10.1038/srep19359/Scientific Reports volume 6}{Sci Rep {\bfseries 6}, 19359 (2016)}.


\bibitem{direct}
LY. Cheng, GH. Yang, Q. Guo et al
\textit{{ Direct measurement of nonlocal entanglement of two-qubit spin quantum states}},
\href{https://doi.org/10.1038/srep19482/Scientific Reports volume 6 }{Sci Rep {\bfseries 6}, 19482 (2016)}.


\bibitem{bar}
R. Lo Franco, G. Compagno
\textit{{ Quantum entanglement of identical particles by standard information-theoretic notions}},
\href{https://doi.org/10.1038/srep20603/Scientific Reports volume 6}{Sci Rep {\bfseries 6}, 20603 (2016)},   [\href{https:// arXiv:quant-ph/1511.03445 }{{\ttfamily
 			arXiv:quant-ph/1511.03445}}].	

\bibitem{L. Amico}
L. Amico, R. Fazio, A. Osterloh, and V. Vedral,
 \textit{{Entanglement in Many-Body Systems}}, 
\href{http://dx.doi.org/10.1103/RevModPhys.80.517}{Rev. Mod. Phys. {\bfseries
  80}, 517 (2008)},  [\href{https://	arXiv:quant-ph/0703044}{{\ttfamily
 	arXiv:quant-ph/0703044}}].
 
 
  \bibitem{to}
K.V. Krutitsky, A. Osterloh , R. Schützhold 
\textit{{ Avalanche of entanglement and correlations at quantum phase transitions}},
\href{https://doi.org/10.1038/s41598-017-03402-8/Scientific Reports volume 7}{Sci Rep {\bfseries 7}, 3634 (2017)},  [\href{https:// arXiv:quant-ph/1607.06616 }{{\ttfamily
 			arXiv:quant-ph/1607.06616 }}].	
 

\bibitem{A. Osterloh}
A. Osterloh, L. Amico, G. Falci, and R. Fazio,
 \textit{{Scaling of Entanglement close to a Quantum Phase Transitions}}, 
\href{https://doi.org/10.1038/416608a}{ Nature {\bfseries
  416}, 608 (2002)},  [\href{https://	arXiv:quant-ph/0202029}{{\ttfamily
 		arXiv:quant-ph/0202029}}].

\bibitem{L.-A. Wu}
 L.-A. Wu, M. S. Sarandy, and D. A. Lidar,
  \textit{{Quantum Phase Transitions and Bipartite Entanglement}}, 
\href{https://doi.org/10.1103/PhysRevLett.93.250404}{ Phys. Rev. Lett. {\bfseries
  93}, 250404 (2004)},  [\href{https:// arXiv:quant-ph/0407056}{{\ttfamily
 			arXiv:quant-ph/0407056}}].
 	
 	
\bibitem{Jaeyoon Cho}
Jaeyoon Cho, Kun Woo Kim
\textit{{Quantum Phase Transition and Entanglement in Topological Quantum Wires}},
\href{https://doi.org/10.1038/s41598-017-02717-w/Scientific Reports volume 7}{Sci Rep {\bfseries 7}, 2745 (2017)},  [\href{https:// arXiv:quant-ph/1706.01637}{{\ttfamily
 			arXiv:quant-ph/1706.01637}}].
 			
 			
\bibitem{ku}
 S.S. Kumar, S. Shankaranarayanan
\textit{{Evidence of quantum phase transition in real-space vacuum entanglement of higher derivative scalar quantum field theories}},
\href{https://doi.org/10.1038/s41598-017-15858-9/Scientific Reports volume 7}{Sci Rep {\bfseries 7}, 15774 (2017)},  [\href{https:// arXiv:cond-mat.stat-mech/1606.05472}{{\ttfamily
 			arXiv:cond-mat.stat-mech/1606.05472}}]. 
 			
 			
 \bibitem{tu}
 M. Gabbrielli, A. Smerzi,  L Pezzè
\textit{{ Multipartite Entanglement at Finite Temperature}},
\href{https://doi.org/10.1038/s41598-018-31761-3/Scientific Reports volume 8}{Sci Rep {\bfseries 8}, 15663 (2018)},  [\href{https:// arXiv:cond-mat.str-el/1805.03139 }{{\ttfamily
 			arXiv:cond-mat.str-el/1805.03139 }}].						
 			
 			
 			
\bibitem{qu}
L. Balazadeh, G. Najarbashi, A. Tavana
\textit{{ Quantum Renormalization of Spin Squeezing in Spin Chains}},
\href{https://doi.org/10.1038/s41598-018-35666-z/Scientific Reports volume 8}{Sci Rep {\bfseries 8}, 17789 (2018)},  [\href{https:// arXiv:cond-mat.str-el/1805.03139 }{{\ttfamily
 			arXiv:cond-mat.str-el/1805.03139 }}].		
 		
 	
 
\bibitem{T. J. Osborne}
T. J. Osborne and M. A. Nielsen, 
\textit{{Entanglement in a simple quantum phase transition}}, 
\href{https://doi.org/10.1103/PhysRevA.66.032110}{ Phys. Rev. A  {\bfseries
  66},  032110 (2002)},  [\href{https:// arXiv:quant-ph/0202162}{{\ttfamily
 				arXiv:quant-ph/0202162}}].

\bibitem{Vidal}
G. Vidal, J. I. Latorre, E. Rico, and A. Kitaev,
\textit{{Entanglement in quantum critical phenomena}}, 
\href{https://doi.org/10.1103/PhysRevLett.90.227902}{Phys. Rev. Lett.  {\bfseries
  90}, 227902 (2003)},  [\href{https://	arXiv:quant-ph/0211074}{{\ttfamily
 					arXiv:quant-ph/0211074}}].


\bibitem{Amit Tribedi}
A. Tribedi and I. Bose,	
\textit{{Quantum critical point and entanglement in a matrix-product ground state}}, 
\href{https://doi.org/10.1103/PhysRevA.75.042304}{Phys. Rev. A.  {\bfseries
  75}, 042304 (2007)}.

\bibitem{W. K. Wootters}
W. K. Wootters, 
\textit{{Entanglement of Formation of an Arbitrary State of Two Qubits}}, 
\href{https://doi.org/10.1103/PhysRevLett.80.2245}{Phys. Rev. Lett.  {\bfseries
  80}, 2245 (1998)},  [\href{https://	arXiv:quant-ph/9709029}{{\ttfamily
 						arXiv:quant-ph/9709029}}].
 	
 	
\bibitem{sahar}
 S. Satoori, S. Mahdavifar, J. Vahedi
\textit{{Entanglement and quantum correlations in the XX spin-1/2 honeycomb lattice}}, 
\href{https://doi.org/10.1038/s41598-022-19945-4/Scientific Reports volume 12}{Sci Rep {\bfseries 12}, 17991 (2022)},  [\href{https:// arXiv:cond-mat.str-el/2204.07708 }{{\ttfamily
 			arXiv:cond-mat.str-el/2204.07708 }}].	
 						 	
	
\bibitem{A. Zheludev}
A. Zheludev, J. M. Tranquada, T. Vogt, and D. J. Buttrey, 
\textit{{Magnetic gap excitations in a one-dimensional mixed spin antiferromagnet Nd2BaNiO5}}, 
\href{https://doi.org/10.1103/PhysRevB.54.7210}{Phys. Rev. B.  {\bfseries 54}, 7210 (1996)}, 


\bibitem{W. Chen}
W. Chen, K. Hida, and B. C. Sanctuary, 
\textit{{Ground State Phase Diagram of S=1 XXZ Chains with Uniaxial Single-Ion-Type Anisotropy}}, 
\href{https://doi.org/10.1103/PhysRevB.67.104401}{Phys. Rev. B.  {\bfseries
  67}, 104401 (2003)},  [\href{https://	arXiv:cond-mat/0209403}{{\ttfamily
 							arXiv:cond-mat/0209403}}].



\bibitem{Zheludev}
A. Zheludev, T. Masuda, I. Tsukada, Y. Uchiyama, K. Uchinokura, P. Boni, and S. H. Lee,

\textit{{Magnetic excitations in coupled Haldane spin chains near the quantum critical point}}, 
\href{https://doi.org/10.1103/PhysRevB.62.8921}{Phys. Rev. B.  {\bfseries
  62}, 8921 (2000)},  


\bibitem{Lou J Z}
J. Z. Lou, T.  Xiang, and Z. B. Su,  
\textit{{Thermodynamics of the bilinear-biquadratic spin one Heisenberg chain}}, 
\href{https://doi.org/10.1103/PhysRevLett.85.2380}{Phys. Rev. Lett.  {\bfseries
  85}, 2380 (2000)},  [\href{arXiv:cond-mat/0003102}{{\ttfamily
 								arXiv:cond-mat/0003102}}].

\bibitem{S.K. Yip}
S. K. Yip, 
\textit{{Dimer state of spin-1 Bosons in an optical lattice}}, 
\href{https://doi.org/10.1103/PhysRevLett.90.250402}{Phys. Rev. Lett.  {\bfseries
  90}, 250402 (2003)},  [\href{arXiv:physics/0306018}{{\ttfamily
 								arXiv: physics/0306018}}].

\bibitem{Fan H}
H. Fan, V.  Korepin and V.  Roychowdhury, 
\textit{{Entanglement in a Valence-Bond-Solid State}}, 
\href{https://doi.org/10.1103/PhysRevLett.93.227203}{Phys. Rev. Lett.  {\bfseries
  93}, 227203 (2004)},  [\href{arXiv:quant-ph/0406067}{{\ttfamily
 								arXiv:quant-ph/0406067}}].

\bibitem{Y. C. Tzeng}
Y. C. Tzeng, H. H. Hung, Y. C. Chen, and M. F. Yang, 
\textit{{Fidelity approach to Gaussian transitions}}, 
\href{https://doi.org/10.1103/PhysRevA.77.062321}{Phys. Rev. A  {\bfseries
  77}, 062321 (2008)},  [\href{arXiv:0804.0537}{{\ttfamily
 									arXiv:0804.0537}}].

\bibitem{Kazuo Hida}
K. Hida and W. Chen, 	
\textit{{Emergence of Long Period Antiferromagnetic Orders from Haldane Phase in S=1 Heisenberg Chains with D-Modulation}}, 
\href{https://doi.org/10.1143/JPSJ.74.2090}{J. Phys. Soc. Jpn  {\bfseries
  74}, 2009 (2005)},  [\href{	arXiv:cond-mat/0504125}{{\ttfamily
 										arXiv:cond-mat/0504125}}].


\bibitem{Haldane}
 F. D. M. Haldane,  
\textit{{Continuum dynamics of the 1-D Heisenberg antiferromagnet: Identification with the O(3) nonlinear sigma model}}, 
\href{https://doi.org/10.1016/0375-9601(83)90631-X}{Physics Letters A  {\bfseries
  93}, 464 (1983)}.
  
  
  \bibitem{Arnesen}
Arnesen M C, Bose S and Vedral V,
\textit{{Natural Thermal and Magnetic Entanglement in 1D Heisenberg Model}}, 
\href{https://doi.org/10.1103/PhysRevLett.87.017901}{Phys. Rev. Lett  {\bfseries
  87}, 017901 (2001)}. [\href{		arXiv:quant-ph/0009060}{{\ttfamily
 													arXiv:quant-ph/0009060}}].

\bibitem{Wang}
Wang X, Fu H, and Solomon A I, \textit{{Thermal entanglement in three-qubit Heisenberg models}}, 
\href{https://doi.org/10.1088/0305-4470/34/50/312}{J. Phys. A: Math. Gen.  {\bfseries
  34},11307 (2001)}. [\href{	arXiv:quant-ph/0105075}{{\ttfamily
 														arXiv:quant-ph/0105075}}].

\bibitem{Schliemann}
Schliemann J et al, \textit{{Quantum Correlations in Two-Fermion Systems}}, 
\href{https://doi.org/10.1103/PhysRevA.64.022303}
{Phys. Rev. A  {\bfseries
  64}, 022303 (2001)}. [\href{		arXiv:quant-ph/0012094}{{\ttfamily
 															arXiv:quant-ph/0012094}}].
 
\bibitem{S. Hill}
S. Hill and W. K. Wootters,
\textit{{Entanglement of a Pair of Quantum Bits}}, 
\href{https://doi.org/10.1103/PhysRevLett.78.5022}{ Phys. Rev. Lett.  {\bfseries
  78}, 5022 (1997)},  [\href{		arXiv:quant-ph/9703041}{{\ttfamily
 											arXiv:quant-ph/9703041}}].




\bibitem{X.-H. Gao}
X.-H. Gao,  A. Sergio, K. Chen, S.-M.  Fei, X.-Q.  Li-Jost,
\textit{{Entanglement of formation and concurrence for mixed states}}, 
\href{https://doi.org/10.1007/s11704-008-0017-8}{ Front. Comput. Sci. China  {\bfseries
  2 2}, 114-128 (2008)}.


\bibitem{G. Vidal}	
G. Vidal and R. F. Werner, 	
\textit{{A computable measure of entanglement}}, 
\href{https://doi.org/10.1103/PhysRevA.65.032314}{Phys. Rev. A  {\bfseries
  65}, 032314 (2002)}. [\href{	arXiv:quant-ph/0102117}{{\ttfamily
 												arXiv:quant-ph/0102117}}].
  
\bibitem{Tao}	
T. Zhou, J. Cui, and G. L. Long, 
\textit{{Measure of nonclassical correlation in coherence-vector representation}}, 
\href{https://doi.org/10.1103/PhysRevA.84.062105}{Phys. Rev. A  {\bfseries
  84}, 062105 (2011)}. 
  
  
  
\bibitem{H}  
 H Bahmani, G Najarbashi and A Tavana, \textit{{Generalized concurrence and quantum phase transition in spin-1 Heisenberg model}}, 
 \href{https://doi.org/10.1088/1402-4896/ab606e}{Phys. Scr {\bfseries
  95}, 055701 (2020)}. 
  
  
  
  \bibitem{H1}  
 H Bahmani, G Najarbashi, B Tarighi and A Tavana, \textit{{Quantum and Classical Thermal Correlations in Spin-1 Heisenberg Chain with Alternating Single-Ion Anisotropy}}, 
 \href{https://doi.org/10.1007/s10909-020-02556-6}{Journal of Low Temperature Physics {\bfseries
  202}, 290-309 (2021)}.
  
  
  
 \bibitem{I. Affleck}
I. Affleck, T. Kennedy, E. H. Lieb, and H. Tasaki, Phys. Rev. Lett. 59, 799 (1987); Commun. Math. Phys. 115, 477 (1988).
\bibitem{M. N. Barber}
M. N. Barber and M. T. Batchelor, Phys. Rev. B 40, 4621 (1989); A. Klumper, ¨ J. Phys. A: Math. Gen. 23, 809 (1990).
  

\bibitem{A. Peres}
A. Peres, \textit{{Separability Criterion for Density Matrices}},
\href{https://doi.org/10.1103/PhysRevLett.77.1413}
{Phys.Rev.Lett {\bfseries
 77}, 1413-1415 (1996)}. [\href{   arXiv:quant-ph/9604005}{{\ttfamily
 																arXiv:quant-ph/9604005}}].

\bibitem{M}
 M. Horodecki, P. Horodecki and R. Horodecki, 
 
 \textit{{Separability of Mixed States: Necessary and Sufficient Conditions}}, 
\href{https://doi.org/10.1016/S0375-9601%2896%2900706-2}
{Phys. Lett. A   {\bfseries
  223}, 1 (1996)}. [\href{arXiv:quant-ph/9605038}{{\ttfamily
 																arXiv:quant-ph/9605038}}].
 
\bibitem{Y Xian}
Y Xian, \textit{{Spontaneous trimerization of spin-1 chains}}, 
\href{https://doi.org/10.1088/0953-8984/5/40/023}{J. Phys.: Candens. Matter  {\bfseries
  5},7489 (1993)}.

\bibitem{Y. Q. Li}
Y. Q. Li, G. Q. Zhu, \textit{{Spontaneous trimerization of spin-1 chains}}, 
\href{https://doi.org/10.1007/s11467-008-0022-2}{Front. Phys. China  {\bfseries
  3(3)},250-257 (2008)}. [\href{	arXiv:quant-ph/0308139}{{\ttfamily
 																	arXiv:quant-ph/0308139}}].
  
\bibitem{T. Zhou}
T. Zhou, J. Cui, and G. L. Long, \textit{{Measure of nonclassical correlation in coherence-vector representation}}, 
\href{https://doi.org/10.1103/PhysRevA.84.062105}{Phys. Rev. A  {\bfseries
  84},062105 (2011)}.	
  \end{thebibliography}
\end{document}